\PassOptionsToPackage{table,dvipsnames}{xcolor}

\documentclass[10pt,twocolumn,letterpaper]{article}

\usepackage[pagenumbers]{cvpr}   

\usepackage{algorithm}
\usepackage{algorithmic}

\usepackage{float}
\usepackage{newfloat}
\usepackage{listings}

\usepackage{array}
\usepackage{multirow}
\usepackage{makecell}

\usepackage[most]{tcolorbox}
\usepackage{wasysym}
\usepackage{fontawesome5}

\definecolor{cvprblue}{rgb}{0.21,0.49,0.74}
\definecolor{oursblue}{RGB}{225,236,253}
\definecolor{avggray}{RGB}{218,221,235}
\definecolor{linkcyan}{RGB}{0,150,190}

\usepackage[
    pagebackref,
    breaklinks,
    colorlinks,
    allcolors=cvprblue
]{hyperref}

\hypersetup{
    pdftitle={
        Training-Free Speech-Centric Omni Understanding with Frozen VLMs
    },
    pdfauthor={
        Ankan Deria, Hanoona Rasheed, Xilin He,
        Fahad Shahbaz Khan, Salman Khan
    },
    pdfsubject={
        Speech-Centric Multimodal Understanding with Frozen
        Vision-Language Models
    },
    pdfkeywords={
        vision-language models,
        speech understanding,
        multimodal understanding,
        training-free inference
    }
}

\floatstyle{ruled}
\newfloat{listing}{tb}{lst}
\floatname{listing}{Listing}

\newcommand{\ourmodel}{TFO}

\newcommand{\rot}[1]{%
    \rotatebox{90}{%
        \hspace{-2pt}%
        \scalebox{0.82}{\textbf{#1}}%
        \hspace{-1pt}%
    }%
}

\def\paperID{00000}
\def\confName{CVPR}
\def\confYear{2026}

\title{
    Training-Free Speech-Centric Omni Understanding with Frozen VLMs
}

\author{
    Ankan Deria
    \quad
    Hanoona Rasheed
    \quad
    Xilin He
    \\[0.2em]
    Fahad Shahbaz Khan
    \quad
    Salman Khan
    \\[0.6em]
    Mohamed bin Zayed University of Artificial Intelligence
    \\
    Abu Dhabi, United Arab Emirates
    \\[0.7em]
    {\small
    \href{https://github.com/mbzuai-oryx/OmniEvalKit}{%
        \textcolor{linkcyan}{%
            \faGithub\;
            \textbf{GitHub:}
            \texttt{github.com/mbzuai-oryx/OmniEvalKit}%
        }%
    }}
    \\[0.3em]
    {\small
    \href{https://mbzuai-oryx.github.io/OmniEvalKit}{%
        \textcolor{linkcyan}{%
            \faGlobe\;
            \textbf{Project Page:}
            \texttt{mbzuai-oryx.github.io/OmniEvalKit}%
        }%
    }}
}

\begin{document}

\maketitle

\begin{abstract}

Audio-visual understanding remains challenging because models must jointly interpret spoken content, visual events, and their temporal relationships. Existing omni models typically introduce dedicated audio encoders and rely on expensive audio-video-text training, tightly coupling omni capability to specific VLM backbones and potentially weakening their existing visual and reasoning abilities. \textit{This raises three questions: whether native omni training is necessary for every new VLM, whether speech-centric omni capability can be added while preserving the original backbone, and where richer acoustic representations remain essential.}

We introduce Training-Free Omni (TFO), a plug-and-play framework that converts a frozen VLM into a speech-centric omni model without architectural modification, or multimodal re-alignment. TFO uses Whisper to extract confidence-filtered, timestamped transcripts and routes them through the VLM's existing language interface, while leaving its visual pathway unchanged. Across matched comparisons with native omni models on 56 benchmarks and 21 languages, TFO is competitive on audio-visual understanding, improves average audio-only performance across all five model settings, and achieves substantial multilingual speech gains. Freezing the VLM also generally preserves stronger image/video understanding, visual grounding, coding, mathematical reasoning, and medical question answering than the corresponding native omni checkpoints. These results show that strong speech-centric omni understanding can often be obtained through modular audio-to-language routing rather than costly backbone-specific training.

\end{abstract}

\section{Introduction}
\label{sec:intro}

Omni models are gaining increasing attention as a step toward models that can understand and generate content across text, images, videos, audio, and speech~\citep{xu2025qwen25omni,xu2025qwen3,cui2026minicpm,miniomni,yu2026salmonn,ye2025omnivinci,audioflamingo}. Native Omni models typically achieve this capability by extending a vision-language model (VLM) with a dedicated audio encoder and aligning its representations with the visual and language representations through large-scale multimodal training. This design enables a single model to process spoken queries, reason jointly over audio and visual content, and produce spoken responses. However, it also tightly couples Omni capability to a particular backbone and training recipe.

This native training presents three main challenges. \textit{First}, it is costly and brittle: as VLM backbones continue to improve, their stronger perception, reasoning, and knowledge capabilities do not automatically transfer to existing Omni models, requiring the audio pathway to be adapted and realigned for each new backbone. \textit{Second}, audio is difficult to integrate reliably because it is temporally dense, often noisy, and must be connected precisely with both spoken content and visual events. Prior studies~\cite{kim2024avhbench,guo2025aligned,jung2026avcd} show that even natively trained audio-visual models may overlook relevant audio, infer sounds from visual cues, or struggle to capture subtle relationships between the two streams. \textit{Third}, modifying and jointly training the backbone can weaken capabilities already present in the original VLM, including image and video understanding, visual grounding, coding, mathematical reasoning, and domain knowledge. Native Omni training must therefore not only acquire audio understanding, but also preserve the mature capabilities of the backbone.

More importantly, Omni tasks require different forms of audio evidence, from recovering spoken content, locating it at the right moment in video, to understanding tone, non-speech sounds, and their fine-grained alignment with visual events. Existing evaluations demonstrate the effectiveness of native Omni models, but they do not determine whether a learned audio pathway is necessary for every task because they lack a matched training-free comparison. This raises a fundamental question: \textit{do we need to train a native Omni model for every new VLM backbone, or can a simple training-free alternative provide comparable speech-centric Omni understanding?}

To investigate this question, we construct \textbf{Training-Free Omni (TFO)}, a simple plug-and-play framework that converts any frozen VLM into a speech-centric Omni model. TFO does not modify the VLM architecture, update its parameters, or require audio-video-text training. Instead, it uses Whisper~\citep{whisper} to recover spoken content and routes it through the VLM's existing language interface. For tasks that require temporal reasoning, TFO retains timestamps that associate spoken segments with the corresponding visual events. Confidence filtering discards unreliable transcripts, and if no segment passes the threshold, \ourmodel{} omits the audio context. TFO therefore adds access to spoken and temporal evidence without introducing a newly trained audio pathway inside the reasoning model.

Our main contribution is a \textit{systematic matched comparison} among native Omni models, their original VLM backbones, and the corresponding training-free conversions across multiple model families and scales. We conduct \textit{extensive evaluation covering 56 benchmarks} spanning audio-visual understanding, audio-only understanding, image and video understanding, visual-grounding, coding and mathematical reasoning, medical question answering, and multilingual speech examined across 21 languages. This design asks \textit{three} complementary questions: \textit{(i)} whether audio routing can recover Omni understanding without native training, \textit{(ii)} whether TFO preserves the capabilities of its VLM backbone that may be weakened during native Omni training, and \textit{(iii)} where richer audio representations remain necessary.

Our study shows that TFO is highly competitive when audio evidence is primarily spoken content, matching or outperforming native Omni models on several audio-visual benchmarks and improving audio-only and multilingual speech understanding across all matched comparisons (Sec.~\ref{sec:main_omni_results}). Across these matched comparisons, TFO generally retains stronger image/video understanding, coding, mathematical reasoning, medical question answering, and visual grounding than the corresponding native Omni models (Sec.~\ref{sec:backbone_preservation}). These results show that native Omni training is not always necessary for strong speech-centric multimodal understanding and may come with measurable capability drift. However, TFO introduces additional ASR latency and remains limited on tasks involving music, environmental sounds, and other non-speech acoustic cues, where transcript-based routing cannot preserve the required acoustic evidence (Sec.~\ref{sec:tradeoffs_limitations}). Together, these findings establish language-level audio routing as a strong control for native Omni training and identify where dedicated acoustic representations remain necessary.
\section{Related Work}
\label{sec:related_work}

\subsection{Native Omni Models and Backbone-Specific Alignment}
Audio-language and Omni models~\citep{xu2025qwen25omni,xu2025qwen3,cui2026minicpm,miniomni,yu2026salmonn,ye2025omnivinci,audioflamingo} commonly extend an LLM or VLM with dedicated acoustic encoders and connectors, followed by audio-text or audio-video-text alignment.
As these audio pathways remain tied to a particular backbone and training recipe, each new VLM generation may require costly multimodal training and cross-modal re-alignment. 
Approaches such as Video-LLaMA~\citep{zhang2023videollama} and Freeze-Omni~\citep{wang2024freezeomni} reduce this cost by freezing parts of the model, but still train the connectors or alignment modules. 
Prior work has used speech transcripts or subtitles as language-side evidence for video understanding, through both training-free model composition and learned modeling~\citep{zeng2022socratic,lei2018tvqa,chen2025livecc}. However, this strategy has not been systematically evaluated as a matched training-free control for native Omni training or used to examine preservation of the original VLM backbone.
In contrast, \ourmodel{} keeps the complete VLM unchanged and studies whether speech-centric Omni understanding can be obtained without training a VLM-side audio pathway.

\subsection{Frozen VLMs and Capability Preservation} 
Modern VLMs provide strong image and video understanding, visual grounding, language reasoning, and domain knowledge~\citep{Qwen2_5VL,yu2026minicpm,NVILA_2025_CVPR,bai2025qwen3vl}. However, existing Omni studies primarily evaluate newly acquired audio capabilities and rarely examine whether multimodal adaptation preserves the original VLM through matched backbone comparisons. Freezing the language backbone has been explored to reduce capability drift~\citep{wang2024freezeomni}, but speech modules and alignment stages are still trained. \ourmodel{} instead freezes the entire VLM and evaluates both sides of the trade-off: the speech-centric Omni capability gained through audio-to-language routing and the visual, reasoning, grounding, and domain-specific capabilities retained from the original backbone.
\begin{figure*}[t]
    \centering
    \includegraphics[width=1.0\linewidth]{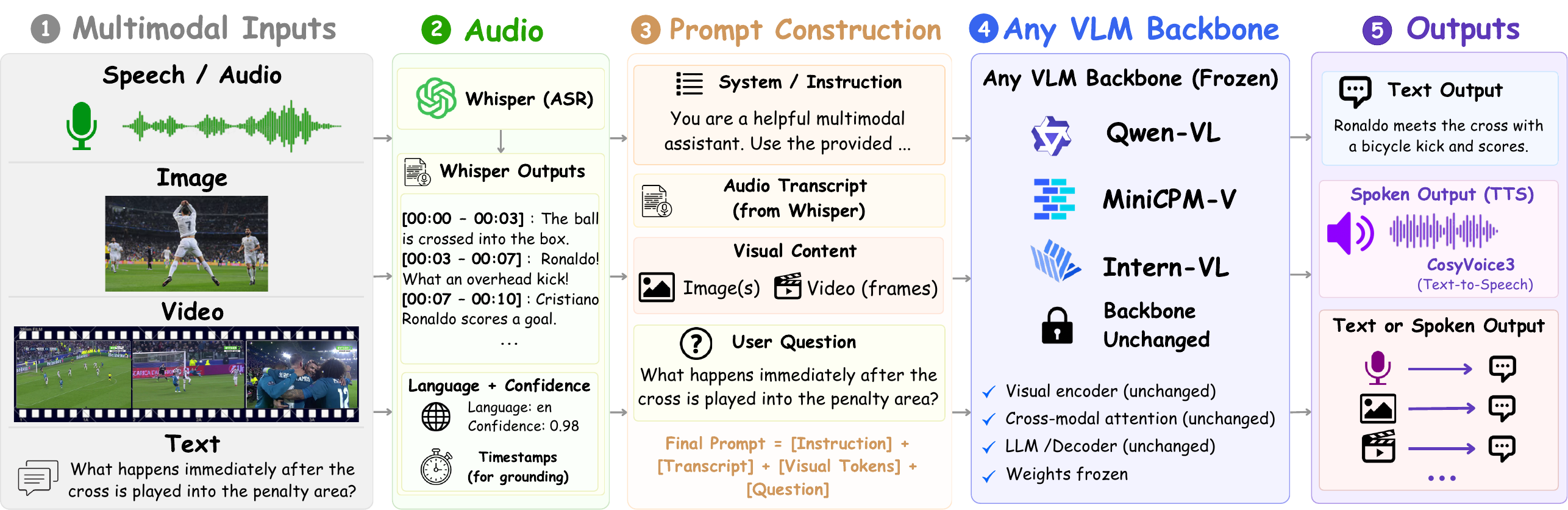}
    \caption{\textbf{Overview of \ourmodel{}.}
    \textbf{(1)} The system accepts speech/audio, image, video, and text inputs.
    \textbf{(2)} Whisper converts audio into transcripts with language, confidence, and timestamp information.
    \textbf{(3)} The transcript, visual content, and user query are combined into a unified prompt.
    \textbf{(4)} A frozen VLM processes the prompt while its visual pathway, architecture, and parameters remain unchanged.
    \textbf{(5)} The model produces a text response, which can optionally be converted to speech using CosyVoice3.
    This modular pipeline supports multimodal inputs and text or spoken outputs without modifying the VLM.}
    \label{fig:pipeline}
\end{figure*}

\section{Methodology}
\label{sec:method}

\subsection{Overview}
\label{sec:overview}
We define Training-Free Omni (\ourmodel{}) as a controlled training-free framework that turns an existing vision-language model (VLM) into a speech-centric omni system without modifying it. The underlying hypothesis is that, for many audio-visual understanding tasks, speech primarily provides linguistic evidence, while the VLM already knows how to reason over language-conditioned visual inputs. Accordingly, we route audio through language rather than introduce a learned audio-token pathway. As shown in Figure~\ref{fig:pipeline}, \ourmodel{} converts audio into text evidence and conditions the frozen VLM through its standard prompt interface.

Let a frozen VLM consist of a visual encoder $E_v$ and language backbone $L_\theta$. Given a visual input $V$, an audio input $A$, and a user query $Q$, \ourmodel{} first converts $A$ into a language-level context $C_A$, then queries the frozen model without updating $\theta$ or introducing any trainable audio module,
\begin{equation}
    y =
    L_\theta\!\left(
        E_v(V),\; P(C_A, Q)
    \right),
    \quad \theta \ \text{frozen}.
    \label{eq:tfo_inference}
\end{equation}
Here, $E_v(V)$ is the original visual representation, and $P(C_A,Q)$ denotes the language-side prompt constructed from the audio context and user query. Thus, \ourmodel{} preserves the visual pathway and changes only the textual context provided to the language backbone.

\subsection{Audio-to-Language Routing}
\label{sec:whisper}

The audio-to-language router constructs the speech evidence used by \ourmodel{}. We instantiate it with Whisper~\citep{whisper}, which operates outside the VLM and decomposes an audio input $A$ into segment-level transcriptions,
\begin{equation}
    \hat{T}(A) =
    \{(w_i, \ell_i, s_i, e_i, c_i)\}_{i=1}^{N},
\end{equation}
where $w_i$ is the transcribed text, $\ell_i$ is the detected language, $s_i$ and $e_i$ are the segment start and end times, and $c_i$ is the transcription confidence. 
This segment-level representation retains the three forms of evidence used by \ourmodel{}: spoken content, temporal boundaries, and transcription reliability.

Not every decoded segment should be passed to the VLM. Silence, background noise, and non-speech regions may produce unreliable or hallucinated transcripts. We therefore retain only segments above a confidence threshold:
\begin{equation}
    T_{\tau}(A) =
    \{(w_i, \ell_i, s_i, e_i)\}_{c_i \geq \tau}.
    \label{eq:confidence_gate}
\end{equation}
The filtered set $T_{\tau}(A)$ is the only audio-derived evidence used by the fusion stage. We set $\tau=0.65$. If no segment passes the threshold, the audio context is omitted and the VLM receives its original visual-language input.

\subsection{\ourmodel{} Multimodal Fusion}
\label{sec:fusion}
Following Eq.~\ref{eq:tfo_inference}, the filtered transcript $T_{\tau}(A)$ is converted into a language-side audio context $C_A$. This context contains the retained speech segments and, for temporal audio-video reasoning, their timestamps. $C_A$ is then inserted into the VLM's standard prompt together with the system instruction, visual input, and user query. Thus, the visual input follows the original VLM pathway, while speech is provided only through the language interface. If $T_{\tau}(A)=\emptyset$, the audio context is omitted. The prompt format and further implementation details are provided in the Supp. (Fig.~\ref{fig:prompt_template} and Sec.~\ref{supp:app_promt_construction}).

For spoken output, we use CosyVoice3~\citep{du2025cosyvoice} to convert the VLM's generated text into speech. 
CosyVoice3 is used only for optional spoken response generation and is not used to construct or modify any evaluation audio. Therefore, its speakers and accents do not affect the reported benchmark results.

\section{Experiments}
\label{sec:experiment}
\begin{table*}[!t]
\centering
\small
\setlength{\tabcolsep}{3.5pt}
\renewcommand{\arraystretch}{1.12}
\resizebox{1.0\textwidth}{!}{%
\begin{tabular}{@{}l
c>{\columncolor{oursblue}}c
c>{\columncolor{oursblue}}c
c>{\columncolor{oursblue}}c
c>{\columncolor{oursblue}}c
c>{\columncolor{oursblue}}c@{}}
\toprule
\multirow{3}{*}{\textbf{Benchmark}}
& \multicolumn{4}{c}{\textbf{Qwen2.5}}
& \multicolumn{2}{c}{\textbf{MiniCPM4.5}}
& \multicolumn{2}{c}{\textbf{VILA}}
& \multicolumn{2}{c}{\textbf{Qwen3}} \\
\cmidrule(lr){2-5}
\cmidrule(lr){6-7}
\cmidrule(lr){8-9}
\cmidrule(lr){10-11}
& \textbf{Omni-3B}
& \textbf{\ourmodel-3B}
& \textbf{Omni-7B}
& \textbf{\ourmodel-7B}
& \textbf{Omni-9B}
& \textbf{\ourmodel-9B}
& \textbf{OmniVinci}
& \textbf{\ourmodel}
& \textbf{Omni-30B}
& \textbf{\ourmodel-30B} \\
\midrule
UnoBench
& 38.4 & 38.0 & 40.5 & 38.4 & 43.6 & 44.3 & 36.6 & 37.5 & 51.3 & 49.1 \\
UnoBench-MC
& 35.0 & 35.1 & 36.7 & 37.1 & 36.4 & 40.2 & 31.6 & 32.8 & 43.4 & 41.5 \\
WorldSense
& 34.2 & 40.3 & 33.9 & 48.4 & 49.5 & 50.6 & 45.2 & 46.8 & 52.3 & 51.9 \\
AV-Odyssey
& 26.3 & 26.8 & 28.8 & 27.3 & 30.2 & 27.6 & 27.2 & 26.1 & 32.6 & 28.7 \\
Video-Holmes
& 34.6 & 42.7 & 39.1 & 43.4 & 56.5 & 59.3 & 43.7 & 44.1 & 52.1 & 50.5 \\
FutureOmni
& 42.9 & 45.9 & 46.3 & 46.5 & 54.3 & 50.5 & 50.1 & 51.5 & 53.9 & 54.0 \\
AVMeme-Full
& 53.0 & 51.3 & 54.2 & 51.6 & 58.0 & 55.0 & 51.2 & 51.9 & 60.7 & 59.2 \\
AVMeme-Main
& 47.0 & 45.0 & 48.4 & 45.8 & 53.0 & 50.0 & 46.0 & 46.2 & 56.0 & 54.8 \\
AVUT-Human
& 55.0 & 63.6 & 59.0 & 65.5 & 73.6 & 67.2 & 63.1 & 63.3 & 70.8 & 72.0 \\
AVUT-Gemini
& 55.5 & 58.9 & 59.5 & 62.4 & 66.2 & 62.6 & 60.5 & 58.5 & 62.4 & 64.8 \\
Daily-Omni
& 53.7 & 60.6 & 57.0 & 61.7 & 79.6 & 68.8 & 58.1 & 59.9 & 68.6 & 68.9 \\
\midrule
\textit{Average}
& 43.2 & 46.2 & 45.8 & 48.0
& 54.6 & 52.4
& 46.7 & 47.1
& 54.9 & 54.1 \\
\bottomrule
\end{tabular}%
}
\caption{
\textbf{AV omni understanding results.}
Evaluation of speech-centric audio-visual understanding across 9 AV omni benchmarks.
We compare \ourmodel{} against native omni counterparts from the same model
family and scale. Results show that language-level audio routing is competitive
with native omni training when spoken evidence and temporal AV contexts are
main.
}
\label{tab:av_benchmark_results}
\end{table*}

Native Omni models typically add dedicated audio pathways and rely on joint audio-video-text alignment. In contrast, \ourmodel{} challenges this design by keeping the VLM frozen and routing speech as language-level evidence. This raises three empirical questions: when does language-level audio routing suffice for Omni understanding, which backbone capabilities does it preserve, and where does native Omni training remain advantageous? \textit{First}, we test whether training-free audio routing can match native Omni models on speech-centric audio-visual and audio-only tasks. \textit{Second}, we assess whether freezing the VLM preserves its image/video understanding, general reasoning, grounding, and domain-specific capabilities. \textit{Third}, we characterize the practical and representational limits of language-level routing, including inference overhead and non-speech acoustic understanding.

\subsection{Evaluation Design}
\label{sec:setup}

We evaluate \ourmodel{} through \textbf{five matched comparisons} across \textbf{four model families}: \textit{(i)} Qwen2.5-VL-Instruct~\citep{Qwen2_5VL} vs.\ Qwen2.5-Omni~\citep{xu2025qwen25omni} at the 3B and 7B scales, \textit{(ii)} MiniCPM-V-4.5~\citep{yu2026minicpm} vs.\ MiniCPM4.5-O~\citep{cui2026minicpm}, \textit{(iii)} NVILA-8B~\citep{NVILA_2025_CVPR} vs.\ OmniVinci~\citep{ye2025omnivinci}, and \textit{(iv)} Qwen3-VL-30B-A3B-Instruct~\citep{bai2025qwen3vl} vs.\ Qwen3-Omni-30B-A3B-Instruct~\citep{xu2025qwen3}. covering audio-visual, image/video, audio-only, and multilingual speech understanding; coding and mathematical reasoning; medical question answering and visual grounding, included as an additional preservation analysis. In total, we cover \textbf{56 benchmark datasets}, including multilingual speech evaluation across 21 CoVoST2 languages. See Supp. Sec.~\ref{supp:app_benchmark_suite} for benchmark details and Supp. Sec.~\ref{supp:app_appendix_evaluation} for additional evaluation details.

\begin{table*}[!t]
\centering
\small
\setlength{\tabcolsep}{3.5pt}
\renewcommand{\arraystretch}{1.12}
\resizebox{1.0\textwidth}{!}{%
\begin{tabular}{@{}l
c>{\columncolor{oursblue}}c
c>{\columncolor{oursblue}}c
c>{\columncolor{oursblue}}c
c>{\columncolor{oursblue}}c
c>{\columncolor{oursblue}}c@{}}
\toprule
\multirow{3}{*}{\textbf{Benchmark}}
& \multicolumn{4}{c}{\textbf{Qwen2.5}}
& \multicolumn{2}{c}{\textbf{MiniCPM4.5}}
& \multicolumn{2}{c}{\textbf{VILA}}
& \multicolumn{2}{c}{\textbf{Qwen3}} \\
\cmidrule(lr){2-5}
\cmidrule(lr){6-7}
\cmidrule(lr){8-9}
\cmidrule(lr){10-11}
& \textbf{Omni-3B}
& \textbf{\ourmodel-3B}
& \textbf{Omni-7B}
& \textbf{\ourmodel-7B}
& \textbf{Omni-9B}
& \textbf{\ourmodel-9B}
& \textbf{OmniVinci}
& \textbf{\ourmodel}
& \textbf{Omni-30B}
& \textbf{\ourmodel-30B} \\
\midrule
Audio Trivia
& 46.3 & 46.4 & 51.4 & 50.8 & 72.4 & 76.7 & 12.9 & 49.0 & 63.5 & 71.1 \\
Audio Web
& 40.3 & 45.6 & 41.6 & 42.0 & 67.8 & 66.9 & 21.8 & 37.2 & 47.6 & 53.1 \\
CoVoST2
& 53.0 & 60.9 & 49.7 & 63.2 & 45.5 & 64.0 & 44.6 & 55.7 & 61.6 & 68.4 \\
FLEURS
& 70.8 & 72.1 & 78.3 & 76.9 & 72.0 & 77.5 & 100.0 & 99.9 & 100.0 & 100.0 \\
LibriSpeech
& 94.9 & 94.6 & 94.0 & 94.1 & 94.1 & 93.0 & 99.1 & 98.2 & 99.1 & 98.6 \\
MELD
& 53.4 & 54.4 & 51.1 & 54.1 & 49.9 & 52.1 & 54.3 & 53.1 & 54.5 & 53.8 \\
MMAR-Bench
& 55.4 & 49.7 & 55.8 & 49.4 & 59.0 & 57.5 & 61.4 & 54.6 & 73.4 & 63.9 \\
Voice-CMMLU
& 55.6 & 58.8 & 55.0 & 59.0 & 57.1 & 55.0 & 32.1 & 48.9 & 65.5 & 66.2 \\
VoiceBench
& 66.0 & 66.2 & 67.2 & 67.5 & 27.5 & 38.5 & 26.7 & 78.0 & 79.8 & 82.9 \\
\midrule
\textit{Average}
& 59.5 & 61.0 & 60.5 & 61.9 & 60.6 & 64.6 & 50.3 & 63.8 & 71.7 & 73.1 \\
\bottomrule
\end{tabular}%
}
\caption{
\textbf{Audio-only understanding.}
Each \ourmodel{} variant is compared with a native Omni counterpart across 9 benchmarks from the same model family and scale. \ourmodel{} achieves a higher overall average across all five model variants, with the strongest gains on speech-dominant tasks, while mixed and non-speech acoustic reasoning remains its main limitation.
}
\label{tab:audio_benchmark_results}
\end{table*}

\subsection{When Is Language-Level Audio Routing Sufficient?}
\label{sec:main_omni_results}

Our \textit{first} finding is that language-level audio routing is competitive with native Omni training on speech-centric tasks, but its effectiveness depends on what the audio signal contributes. When the relevant evidence is primarily spoken content, \ourmodel{} often matches or surpasses native Omni models by converting speech into text. When the task requires richer acoustic perception or tightly learned audio-visual alignment, native Omni training can retain an advantage.

Across \textbf{9 audio-visual understanding benchmarks}, Table~\ref{tab:av_benchmark_results} confirms this pattern. \ourmodel{} improves the Qwen2.5 average by \textbf{+3.0} points for 3B and \textbf{+2.2} for 7B, while VILA gains \textbf{+0.4}; MiniCPM4.5 and Qwen3 decrease by \textbf{2.2} and \textbf{0.8} points, respectively. The strongest gains across both Qwen2.5 scales occur in speech-conditioned and temporal video reasoning: WorldSense improves by \textbf{+6.1}/\textbf{+14.5}, Video-Holmes by \textbf{+8.1}/\textbf{+4.3}, AVUT-Human by \textbf{+8.6}/\textbf{+6.5}, and Daily-Omni by \textbf{+6.9}/\textbf{+4.7}. These results show that timestamped audio routing is particularly effective when spoken evidence must align with visual events over time.

\begin{table*}[!t]
\centering
\setlength{\tabcolsep}{2.2pt}
\renewcommand{\arraystretch}{1.12}
\resizebox{1.0\textwidth}{!}{%
\begin{tabular}{@{}lcccccccccccccccccccccc@{}}
\toprule
\textbf{Model}
& \textbf{ar}
& \textbf{ca}
& \textbf{cy}
& \textbf{de}
& \textbf{es}
& \textbf{et}
& \textbf{fa}
& \textbf{fr}
& \textbf{id}
& \textbf{it}
& \textbf{ja}
& \textbf{lv}
& \textbf{mn}
& \textbf{nl}
& \textbf{pt}
& \textbf{ru}
& \textbf{sl}
& \textbf{sv}
& \textbf{ta}
& \textbf{tr}
& \textbf{zh}
& \textbf{Avg.} \\
\midrule

Qwen2.5-Omni-3B
& 70.2 & 63.2 & 2.5 & 89.0 & 92.5 & 2.9 & 4.4 & 90.7 & 82.8 & 90.5 & 66.7 & 4.3 & 0.7 & 86.2 & 93.4 & 93.2 & 18.3 & 34.9 & 3.2 & 45.6 & 78.4 & 53.0 \\
\rowcolor{oursblue}
Qwen2.5-VL-\ourmodel-3B
& 58.8 & 69.0 & 15.7 & 87.5 & 89.1 & 45.8 & 35.9 & 87.5 & 84.8 & 88.6 & 66.2 & 29.2 & 5.2 & 85.3 & 94.0 & 92.0 & 43.9 & 64.5 & 5.2 & 65.8 & 64.1 & \textbf{60.9} \\

Qwen2.5-Omni-7B
& 69.6 & 60.8 & 3.2 & 88.8 & 93.1 & 2.1 & 2.9 & 90.4 & 85.0 & 90.8 & 40.8 & 4.0 & 1.1 & 86.7 & 93.7 & 93.4 & 20.0 & 14.0 & 1.6 & 23.9 & 76.9 & 49.7 \\
\rowcolor{oursblue}
Qwen2.5-VL-\ourmodel-7B
& 60.4 & 72.2 & 22.6 & 87.5 & 89.4 & 50.6 & 39.9 & 88.2 & 86.8 & 89.6 & 65.9 & 36.7 & 5.1 & 85.5 & 93.6 & 92.0 & 48.6 & 68.9 & 9.0 & 70.5 & 64.2 & \textbf{63.2} \\

\midrule
MiniCPM4.5-O-9B
& 34.6 & 51.1 & 4.8 & 74.0 & 83.9 & 8.9 & 16.4 & 77.8 & 65.4 & 76.8 & 39.0 & 4.0 & 1.1 & 68.2 & 79.2 & 84.3 & 19.7 & 35.9 & 5.1 & 53.0 & 73.4 & 45.6 \\
\rowcolor{oursblue}
MiniCPM4.5-V-\ourmodel-9B
& 60.9 & 72.3 & 26.8 & 86.8 & 89.6 & 52.5 & 40.3 & 86.2 & 87.9 & 87.3 & 68.1 & 39.4 & 5.1 & 86.2 & 92.8 & 92.0 & 48.3 & 71.7 & 14.0 & 69.1 & 65.8 & \textbf{64.0} \\

\midrule
OmniVinci-9B
& 40.2 & 63.1 & 2.8 & 84.0 & 90.0 & 2.3 & 3.3 & 89.3 & 46.6 & 87.5 & 58.2 & 3.5 & 0.9 & 76.2 & 90.2 & 89.1 & 14.4 & 14.7 & 1.9 & 13.9 & 64.7 & 44.6 \\
\rowcolor{oursblue}
NVILA-8B-VL-\ourmodel
& 45.1 & 63.4 & 12.6 & 86.2 & 89.6 & 28.7 & 20.5 & 86.6 & 82.6 & 86.1 & 65.4 & 23.5 & 3.8 & 83.3 & 91.2 & 90.0 & 35.3 & 58.2 & 4.3 & 52.4 & 60.4 & \textbf{55.7} \\

\midrule
Qwen3-Omni-30B-A3B
& 79.0 & 67.5 & 3.3 & 91.0 & 92.5 & 17.8 & 37.6 & 92.0 & 92.8 & 91.0 & 75.9 & 7.6 & 0.7 & 91.4 & 95.2 & 93.6 & 38.6 & 68.7 & 7.1 & 72.9 & 76.9 & 61.6 \\
\rowcolor{oursblue}
Qwen3-VL-30B-A3B-\ourmodel
& 65.6 & 75.6 & 32.5 & 88.4 & 90.5 & 67.6 & 47.0 & 88.5 & 91.2 & 89.6 & 72.7 & 49.1 & 4.1 & 87.5 & 94.7 & 94.8 & 55.0 & 73.4 & 21.1 & 76.2 & 70.7 & \textbf{68.4} \\

\bottomrule
\end{tabular}%
}
\caption{
\textbf{Multilingual speech understanding.}
Across 21 CoVoST2 languages, \ourmodel{} consistently improves across all 5 model comparisons, showing multilingual speech capability can be effectively transferred from the ASR front-end to a frozen VLM.}
\label{tab:multilingual}
\end{table*}


\begin{table*}[!t]
\centering
\small
\setlength{\tabcolsep}{4.0pt}
\renewcommand{\arraystretch}{1.12}
\resizebox{1.0\textwidth}{!}{%
\begin{tabular}{@{}l
cccccccc>{\columncolor{oursblue}}c
ccccccc>{\columncolor{oursblue}}c@{}}
\toprule
& \multicolumn{9}{c}{\textbf{Image Benchmarks}}
& \multicolumn{8}{c}{\textbf{Video Benchmarks}} \\
\cmidrule(lr){2-10}
\cmidrule(lr){11-18}

\textbf{Model}
& \rot{ChartQA}
& \rot{DocVQA}
& \rot{InfoVQA}
& \rot{OCRBench}
& \rot{TextVQA}
& \rot{MMBench}
& \rot{MMStar}
& \rot{MME}
& \rot{Avg. (Image)}
& \rot{LongVideo}
& \rot{LVBench}
& \rot{MotionBench}
& \rot{VideoMME}
& \rot{VideoMME-S}
& \rot{EgoSchema}
& \rot{MVBench}
& \rot{Avg. (Video)} \\
\midrule

Qwen2.5-Omni-3B
& 82.8 & 92.7 & 69.4 & 77.5 & 79.7
& 84.2 & 54.9 & 85.4 & 78.3
& 51.9 & 36.8 & 53.7 & 59.3 & 65.2
& 56.0 & 62.6 & 55.1 \\

\rowcolor{oursblue}
Qwen2.5-VL-\ourmodel-3B
& 84.2 & 93.9 & 77.1 & 79.7 & 79.3
& 84.6 & 56.5 & 85.7 & 80.1
& 54.2 & 43.3 & 52.1 & 61.5 & 71.7
& 66.6 & 62.1 & 58.8 \\

Qwen2.5-Omni-7B
& 85.3 & 95.2 & 79.4 & 85.7 & 84.4
& 87.9 & 62.8 & 87.2 & 83.5
& 51.9 & 42.8 & 52.1 & 60.9 & 70.0
& 61.6 & 63.6 & 57.6 \\

\rowcolor{oursblue}
Qwen2.5-VL-\ourmodel-7B
& 87.3 & 95.7 & 82.6 & 86.4 & 84.9
& 88.2 & 62.8 & 87.7 & 84.5
& 56.0 & 45.3 & 52.3 & 71.4 & 75.8
& 65.2 & 63.8 & 61.4 \\

\midrule

MiniCPM4.5-O-9B
& 85.8 & 94.9 & 72.8 & 87.1 & 82.5
& 89.6 & 64.9 & 89.8 & 83.4
& 65.8 & 50.9 & 59.7 & 67.5 & 83.9
& -- & 60.4 & 64.7 \\

\rowcolor{oursblue}
MiniCPM4.5-V-\ourmodel-9B
& 86.2 & 95.2 & 76.8 & 88.7 & 82.3
& 89.3 & 64.9 & 89.6 & 84.1
& 64.0 & 50.5 & 60.1 & 67.9 & 83.6
& -- & 60.5 & 64.4 \\

\midrule

OmniVinci-9B
& 75.3 & 86.7 & 59.6 & 77.1 & 27.4
& 90.1 & 59.8 & 80.4 & 69.6
& 59.0 & 40.7 & 56.7 & 68.1 & 76.9
& 67.2 & 67.3 & 62.3 \\

\rowcolor{oursblue}
NVILA-8B-VL-\ourmodel
& 72.3 & 77.3 & 39.1 & 65.0 & 77.9
& 89.9 & 57.4 & 80.6 & 69.9
& 60.2 & 45.5 & 58.5 & 72.8 & 78.2
& 68.6 & 68.6 & 64.6 \\

\midrule

Qwen3-Omni-30B-A3B
& 80.4 & 94.8 & 80.8 & 85.8 & 91.2
& 90.8 & 70.3 & 89.7 & 85.5
& 64.9 & 44.6 & 64.2 & 72.7 & 80.6
& 64.8 & 68.9 & 65.8 \\

\rowcolor{oursblue}
Qwen3-VL-30B-A3B-\ourmodel
& 81.2 & 95.8 & 87.8 & 92.3 & 92.9
& 91.6 & 71.9 & 89.7 & 87.9
& 66.7 & 48.8 & 64.5 & 78.1 & 80.7
& 71.6 & 70.6 & 68.7 \\

\bottomrule
\end{tabular}
}
\caption{
\textbf{Image and video understanding results.}
Across eight image and up to six video benchmarks, \ourmodel{} generally matches or outperforms the corresponding native Omni models, showing strong preservation of the original VLM's visual capabilities.
}
\label{tab:iv_benchmark_results}
\end{table*}

Beyond AV reasoning, Table~\ref{tab:audio_benchmark_results} evaluates \ourmodel{} across \textbf{9 audio-only understanding benchmarks}. The average improves across all five settings by \textbf{+1.5}, \textbf{+1.4}, \textbf{+4.0}, \textbf{+13.5}, and \textbf{+1.4} points. The largest gain occurs on VILA (\textbf{50.3$\rightarrow$63.8}), while CoVoST2 and VoiceBench improve across every model family, including VoiceBench gains of \textbf{+51.3} on VILA and \textbf{+11.0} on MiniCPM4.5. This consistent improvement on speech-dominant tasks supports the effectiveness of routing spoken content through language. In contrast, all variants decline on MMAR-Bench by \textbf{1.5--9.5} points, highlighting the limitation of transcript-based routing for music, sound events, and broader non-speech acoustic reasoning.

The modular design provides a further advantage in multilingual settings. Across \textbf{21 languages}, Table~\ref{tab:multilingual} shows consistent improvements in \textbf{multilingual speech understanding} on CoVoST2. \ourmodel{} improves the average across all five settings by \textbf{+7.9}, \textbf{+13.5}, \textbf{+18.4}, \textbf{+11.1}, and \textbf{+6.8} points, with the largest gain on MiniCPM4.5 (\textbf{45.6$\rightarrow$64.0}). The largest improvements occur where native Omni models are weak, including Estonian (\textbf{+43.6}), Latvian (\textbf{+35.4}), and Swedish (\textbf{+35.8}) on MiniCPM4.5, and Swedish (\textbf{+54.9}) and Turkish (\textbf{+46.6}) on Qwen2.5-7B. The consistent gains across model families demonstrate that a strong ASR front-end can transfer multilingual speech coverage to a frozen VLM without backbone-specific audio alignment.

\subsection{Does \ourmodel{} Preserve the VLM Backbone?}
\label{sec:backbone_preservation}
Our \textit{second} finding is that training-free conversion retains the capabilities of the original VLM more reliably than native Omni training. We evaluate image/video understanding, coding and mathematical reasoning, medical question answering, and visual grounding, where audio-to-language routing offers no direct advantage. These tasks therefore isolate retention of the backbone's visual, reasoning, grounding, and domain-specific capabilities.

We first examine \textbf{image/video understanding} in Table~\ref{tab:iv_benchmark_results}, covering \textbf{eight image} and \textbf{six video} benchmarks. Compared with native Omni counterparts, \ourmodel{} achieves higher image averages across all five settings, with gains of \textbf{+1.8}, \textbf{+1.0}, \textbf{+0.7}, \textbf{+0.3}, and \textbf{+2.4} points. Except for NVILA, the clearest image gains appear on document, chart, and OCR-intensive tasks. Video preservation is more consistent, with higher averages in four settings by \textbf{+3.7}, \textbf{+3.8}, \textbf{+2.3}, and \textbf{+2.9} points, while MiniCPM4.5 remains nearly unchanged. VideoMME improves across every model family by \textbf{+0.4} to \textbf{+10.5} points, and recurring gains on LVBench and VideoMME indicate that \ourmodel{} retains long-context video reasoning while adding speech-centric Omni capability.

\begin{table*}[!t]
\centering
\small
\setlength{\tabcolsep}{3.5pt}
\renewcommand{\arraystretch}{1.12}
\resizebox{1.0\textwidth}{!}{%
\begin{tabular}{@{}lccccccccc@{}}
\toprule
\multirow{4}{*}{\textbf{Model}}
& \multicolumn{3}{c}{\textbf{Coding}}
& \multicolumn{5}{c}{\textbf{Math}}
& \textbf{Avg.} \\
\cmidrule(lr){2-4}
\cmidrule(lr){5-9}
& \textbf{MBPP}
& \textbf{MBPP Sanit}
& \textbf{HumEval}
& \textbf{GSM8K}
& \textbf{MATH500}
& \textbf{MathVerse}
& \textbf{MathVista}
& \textbf{VideoMath}
& \\
\midrule

Qwen2.5-Omni-3B
& 51.4 & 58.8 & 59.1 & 76.1 & 45.4 & 32.1 & 59.5 & 24.0 & 50.8 \\
\rowcolor{oursblue}
Qwen2.5-VL-\ourmodel-3B
& 56.3 & 65.3 & 61.6 & 77.9 & 44.6 & 47.6 & 62.3 & 25.0 & \textbf{55.1} \\

Qwen2.5-Omni-7B
& 61.1 & 68.4 & 70.1 & 84.2 & 55.2 & 40.2 & 68.0 & 23.6 & 58.9 \\
\rowcolor{oursblue}
Qwen2.5-VL-\ourmodel-7B
& 65.8 & 71.9 & 72.6 & 84.5 & 58.0 & 49.2 & 68.2 & 24.5 & \textbf{61.8} \\

\midrule
OmniVinci-9B
& 53.5 & 55.5 & 38.4 & 64.2 & 36.8 & 29.4 & 69.1 & 24.1 & \textbf{46.4} \\
\rowcolor{oursblue}
NVILA-8B-VL-\ourmodel
& 54.4 & 56.0 & 39.0 & 62.8 & 26.8 & 25.1 & 58.0 & 26.4 & 43.6 \\

\midrule
Qwen3-Omni-30B-A3B
& 95.6 & 97.0 & 90.1 & 96.2 & 83.6 & 38.2 & 50.1 & 34.0 & 73.1 \\
\rowcolor{oursblue}
Qwen3-VL-30B-A3B-\ourmodel
& 98.5 & 97.9 & 92.2 & 96.2 & 84.0 & 40.0 & 61.4 & 34.5 & \textbf{75.6} \\

\bottomrule
\end{tabular}
}
\caption{
\textbf{Coding and mathematical reasoning across seven benchmarks.}
\ourmodel{} achieves a higher overall average in three of the four model comparisons and consistently outperforms its corresponding native Omni counterpart on both coding benchmarks, while performance on mathematical reasoning is more model-dependent.}
\label{tab:coding_math}
\end{table*}


\begin{table*}[!t]
\centering
\small
\setlength{\tabcolsep}{3.2pt}
\renewcommand{\arraystretch}{1.12}
\resizebox{1.0\textwidth}{!}{%
\begin{tabular}{@{}l
c>{\columncolor{oursblue}}c
c>{\columncolor{oursblue}}c
c>{\columncolor{oursblue}}c
c>{\columncolor{oursblue}}c
c>{\columncolor{oursblue}}c@{}}
\toprule
\textbf{Benchmark}
& \multicolumn{4}{c}{\textbf{Qwen2.5}}
& \multicolumn{2}{c}{\textbf{MiniCPM4.5}}
& \multicolumn{2}{c}{\textbf{VILA}}
& \multicolumn{2}{c}{\textbf{Qwen3}} \\
\cmidrule(lr){2-5}
\cmidrule(lr){6-7}
\cmidrule(lr){8-9}
\cmidrule(lr){10-11}
& \textbf{Omni-3B}
& \textbf{VL-\ourmodel-3B}
& \textbf{Omni-7B}
& \textbf{VL-\ourmodel-7B}
& \textbf{O-9B}
& \textbf{V-\ourmodel-9B}
& \textbf{OmniVinci}
& \textbf{VL-\ourmodel}
& \textbf{Omni-30B}
& \textbf{VL-30B-\ourmodel} \\
\midrule
MMMU-Med-test
& 44.5 & 45.0 & 50.1 & 51.0 & 56.2 & 58.3 & 46.8 & 46.2 & 63.4 & 65.8 \\
MMMU-Med-val
& 45.3 & 51.3 & 48.0 & 54.0 & 56.7 & 58.7 & 43.3 & 45.3 & 66.6 & 69.3 \\
MedFrameQA
& 44.0 & 45.4 & 45.6 & 47.9 & 50.3 & 54.0 & 53.0 & 53.0 & 40.8 & 50.6 \\
MedMCQA
& 49.4 & 50.9 & 53.8 & 55.9 & 53.7 & 54.4 & 49.8 & 51.9 & 67.4 & 68.5 \\
MedQA-USMLE
& 49.5 & 51.5 & 57.5 & 58.4 & 61.8 & 64.6 & 52.7 & 53.7 & 76.5 & 77.2 \\
MedXQA-MM
& 19.5 & 21.2 & 20.8 & 22.3 & 23.7 & 24.4 & 22.2 & 22.7 & 30.9 & 28.2 \\
MedXQA-Text
& 10.7 & 10.9 & 11.8 & 12.9 & 14.5 & 14.9 & 12.2 & 12.7 & 17.8 & 17.8 \\
Medbullets-op4
& 43.1 & 43.9 & 46.4 & 48.4 & 55.8 & 56.5 & 42.5 & 44.2 & 63.6 & 63.6 \\
Medbullets-op5
& 37.0 & 37.5 & 40.3 & 37.0 & 45.8 & 46.5 & 37.0 & 37.4 & 55.2 & 56.2 \\
OmniMedVQA
& 65.4 & 66.7 & 62.6 & 63.4 & 76.4 & 78.5 & 72.1 & 73.8 & 77.3 & 80.1 \\
PATH-VQA
& 32.2 & 32.9 & 34.4 & 34.1 & 36.8 & 37.6 & 58.4 & 54.2 & 36.8 & 37.8 \\
PMC-VQA
& 47.5 & 50.7 & 50.6 & 51.2 & 58.4 & 64.0 & 52.9 & 55.5 & 50.3 & 52.9 \\
PubMedQA
& 70.0 & 73.6 & 74.6 & 76.0 & 74.0 & 75.0 & 76.2 & 75.8 & 76.4 & 77.0 \\
SGPQ
& 24.8 & 25.0 & 26.2 & 27.8 & 31.0 & 31.0 & 26.4 & 26.5 & 45.0 & 44.7 \\
\midrule
\textit{Average}
& 41.6 & 43.3 & 44.5 & 45.7
& 49.7 & 51.3
& 46.1 & 46.6
& 54.9 & 56.4 \\
\bottomrule
\end{tabular}%
}
\caption{
\textbf{Medical question answering across 12 benchmarks.}
\ourmodel{} improves the average score across all five model families, showing stronger preservation of domain-specific knowledge and clinical reasoning than the corresponding native Omni models.
}
\label{tab:all_medical_benchmark}
\end{table*}

We next examine whether the same retention holds for \textbf{coding and mathematical reasoning}. Table~\ref{tab:coding_math} shows that \ourmodel{} achieves higher overall averages in three of four settings, with gains of \textbf{+4.3}, \textbf{+2.9}, and \textbf{+2.5} points on Qwen2.5-3B, Qwen2.5-7B, and Qwen3. Every \ourmodel{} variant outperforms its native Omni counterpart on MBPP, MBPP Sanitized, and HumanEval, showing consistent preservation of coding ability. For the Qwen models, text-based mathematics remains stable, while larger gains appear in visual reasoning, including \textbf{+15.5} and \textbf{+9.0} on MathVerse for 3B and 7B, and \textbf{+11.3} on MathVista for Qwen3. NVILA is the main exception, with a \textbf{-2.8} point average decline driven primarily by lower mathematical reasoning scores. Overall, the results suggest that avoiding native Omni re-alignment generally reduces drift from reasoning capabilities already learned by the backbone.

Table~\ref{tab:all_medical_benchmark} extends this analysis to domain-specific knowledge and multimodal clinical reasoning across \textbf{12 medical question answering benchmarks}. Relative to native omni counterparts, \ourmodel{} achieves higher averages across all five settings, with gains of \textbf{+1.7}, \textbf{+1.2}, \textbf{+1.6}, \textbf{+0.5}, and \textbf{+1.5} points. The improvements are broad, with MedMCQA, MedQA-USMLE, and OmniMedVQA increasing in every setting. Larger gains include \textbf{+9.8} on MedFrameQA for Qwen3, \textbf{+5.6} on PMC-VQA for MiniCPM4.5, and \textbf{+6.0} on MMMU-Med-val for both Qwen2.5 scales. These results further indicate that freezing the backbone also better retains specialized knowledge and domain-specific reasoning.

Finally, Table~\ref{tab:grounding_results} evaluates \textbf{visual grounding} as an additional preservation test, which receives no direct benefit from audio-to-language routing.. \ourmodel{} improves PixMo-Count across all five settings, including gains of \textbf{+15.5} on VILA and \textbf{+11.2} on Qwen3, while PixMo-Point error decreases in four settings and remains tied on VILA. PointArena also improves in four settings and ties on VILA, whereas RefCOCO is more model-dependent, with a particularly large gain for Qwen3. These results indicate that the frozen visual pathway retains its grounding behavior under speech-centric Omni conversion. Together, these results show that freezing the VLM broadly preserves its visual, reasoning, grounding, and domain-specific capabilities.

\subsection{Practical Trade-offs and Limitations}
\label{sec:tradeoffs_limitations}

Our \textit{third} finding is that training-free routing introduces two distinct trade-offs: it shifts cost from training to inference, and it cannot represent acoustic evidence that is absent from speech transcripts. Table~\ref{tab:tradeoffs} quantifies the first trade-off. \ourmodel{} remains comparable in parameter count to native Omni models, with smaller Qwen2.5 variants and only a marginal increase for MiniCPM4.5. Its main practical cost is sequential ASR inference. On AVMeme, total latency increases from approximately 0.7--2.4\,s for native Omni models to 1.3--3.1\,s for \ourmodel{}. This overhead can be amortized when several questions share the same audio-video input because the transcript is generated once and reused. This additional inference cost accompanies the central advantage of \ourmodel{}: eliminating backbone-specific audio training and multimodal re-alignment while adding speech-centric Omni capability to frozen VLMs through a modular front-end.

\begin{table}[!t]
\centering
\small
\setlength{\tabcolsep}{3.2pt}
\renewcommand{\arraystretch}{1.08}
\resizebox{0.94\columnwidth}{!}{%
\begin{tabular}{@{}lcccc@{}}
\toprule
\makecell{\textbf{Model}\\\textbf{}}
& \makecell{\textbf{PixMo}\\\textbf{Count}}
& \makecell{\textbf{PixMo}\\\textbf{Point} $\downarrow$}
& \makecell{\textbf{RefCOCO}\\\textbf{test}}
& \makecell{\textbf{Point}\\\textbf{Arena}} \\
\midrule
Qwen2.5-Omni-3B
& 52.3 & 1.26 & 80.2 & 2.1 \\
\rowcolor{oursblue}
Qwen2.5-VL-\ourmodel-3B
& 56.4 & 1.22 & 80.5 & 2.5 \\
Qwen2.5-Omni-7B
& 62.1 & 1.25 & 80.9 & 9.2 \\
\rowcolor{oursblue}
Qwen2.5-VL-\ourmodel-7B
& 62.3 & 1.18 & 81.7 & 9.7 \\
\midrule
MiniCPM4.5-O-9B
& 59.1 & 20.30 & 6.2 & 17.4 \\
\rowcolor{oursblue}
MiniCPM4.5-V-\ourmodel-9B
& 66.3 & 19.07 & 4.4 & 18.4 \\
\midrule
OmniVinci-9B
& 43.6 & 1.31 & 0.0 & 0.0 \\
\rowcolor{oursblue}
NVILA-8B-VL-\ourmodel
& 59.1 & 1.31 & 0.0 & 0.0 \\
\midrule
Qwen3-Omni-30B-A3B
& 56.4 & 1.28 & 6.1 & 0.0 \\
\rowcolor{oursblue}
Qwen3-VL-30B-A3B-\ourmodel
& 67.6 & 1.21 & 85.0 & 7.1 \\
\bottomrule
\end{tabular}%
}
\caption{
\textbf{Visual grounding results.}
\ourmodel{} achieves stronger performance on most counting, pointing, and referring-expression metrics. PixMo-Count and PointArena report accuracy, PixMo-Point reports normalized Euclidean distance ($\downarrow$), and RefCOCO reports mean IoU.
}
\label{tab:grounding_results}
\end{table}

The more fundamental limitation is representational. We use AVHBench to separate cases that require verifying the audio from those that require verifying the video. In \textit{AV Matching}, the model must determine whether the observed sound matches the visual event. In \textit{Video-Driven Audio Hallucination} (V$\rightarrow$A), it must verify whether a sound suggested by the video is actually present in the audio. Both settings require access to non-speech acoustic evidence, which is lost when audio is reduced to a transcript. In contrast, \textit{Audio-Driven Video Hallucination} (A$\rightarrow$V) requires verifying whether an object suggested by the audio is actually visible. Since \ourmodel{} preserves the original visual pathway, it performs better than native omni models in this setting across all four model families. These results clearly separate the scope of language-level routing: it supports faithful reasoning over visual evidence, but cannot verify sounds that are absent from the transcript.

To examine whether this limitation can be addressed without changing the routing paradigm, we further append textual predictions from auxiliary audio models. Further analysis is provided in Supp. Sec. \ref{supp:app_additional_ablations}; Tables~\ref{tab:aba_av_benchmark_results} and~\ref{tab:aba_audio_benchmark_results} show that MELLOW and SenseVoice provide isolated gains, but neither consistently improves over Whisper alone. This suggests that richer acoustic evidence cannot be reliably recovered through textual audio outputs.

\begin{table}[!t]
\centering
\scriptsize
\setlength{\tabcolsep}{2.2pt}
\renewcommand{\arraystretch}{1.02}
\resizebox{0.96\columnwidth}{!}{%
\begin{tabular}{@{}lccc|ccc@{}}
\toprule
\textbf{Model}
& \multicolumn{3}{c|}{\textbf{AVHBench}}
& \multicolumn{3}{c}{\textbf{AVMeme}} \\
\cmidrule(lr){2-4}
\cmidrule(l){5-7}
& \shortstack{\textbf{AV}\\\textbf{Match} $\uparrow$}
& \shortstack{\textbf{V$\rightarrow$A}\\\textbf{Hall.} $\uparrow$}
& \shortstack{\textbf{A$\rightarrow$V}\\\textbf{Hall.} $\uparrow$}
& \shortstack{\textbf{Whisper}\\\textbf{Time (s)}}
& \shortstack{\textbf{VLM}\\\textbf{Time (s)}}
& \shortstack{\textbf{Params}\\\textbf{(B)}} \\
\midrule

MiniCPM4.5-O-9B
& \textbf{74.41} & \textbf{79.43} & 82.57
& -- & 0.82 & \textbf{9.4} \\

\rowcolor{oursblue}
MiniCPM4.5-TFO-9B
& 49.25 & 70.09 & \textbf{84.60}
& 0.62 & 0.69 & 9.5 \\
\midrule

Qwen2.5-O-3B
& \textbf{61.41} & \textbf{76.16} & 77.08
& -- & 0.66 & 5.5 \\

\rowcolor{oursblue}
Qwen2.5-TFO-3B
& 58.48 & 67.77 & \textbf{79.49}
& 0.73 & 0.61 & \textbf{4.6} \\
\midrule

Qwen2.5-O-7B
& \textbf{72.39} & \textbf{79.78} & 79.67
& -- & 0.77 & 10.7 \\

\rowcolor{oursblue}
Qwen2.5-TFO-7B
& 56.29 & 63.36 & \textbf{81.25}
& 0.73 & 0.77 & \textbf{9.1} \\
\midrule

Qwen3-O-30B-A3B
& \textbf{61.57} & \textbf{75.72} & 77.64
& -- & 2.39 & 31.72 \\

\rowcolor{oursblue}
Qwen3-TFO-30B-A3B
& 61.25 & 68.30 & \textbf{86.80}
& 0.74 & 2.34 & 31.88 \\
\bottomrule
\end{tabular}%
}
\caption{Practical trade-offs of \ourmodel{}. AVHBench evaluates audio-visual matching and cross-modal hallucination, while AVMeme reports average Whisper and VLM inference times and parameter count. V$\rightarrow$A and A$\rightarrow$V denote video-driven audio and audio-driven video hallucination, respectively.}
\label{tab:tradeoffs}
\end{table}

\begin{table}[!t]
\centering
\scriptsize
\setlength{\tabcolsep}{2.2pt}
\renewcommand{\arraystretch}{0.9}
\resizebox{0.95\columnwidth}{!}{%
\begin{tabular}{@{}lcccccc@{}}
\toprule
\textbf{Method}
& \rot{UnoBench}
& \rot{WorldSense}
& \rot{V-Holmes}
& \rot{AVUT-H}
& \rot{AVUT-G}
& \rot{Daily-Omni} \\
\midrule

Qwen2.5-Omni-3B
& 38.4 & 34.2 & 34.6 & 55.0 & 55.5 & 53.7 \\
\midrule

Qwen2.5-VL-3B
& 24.4 & 39.7 & 39.9 & 52.7 & 42.9 & 47.2 \\

\rowcolor{oursblue}
\hspace{0.4em}+ Whisper
& 38.0 & 40.3 & 42.7 & 63.6 & 58.9 & 60.6 \\

\hspace{0.4em}+ Whisper (W/o TS)
& 38.0 & 40.1 & 42.5 & 62.9 & 56.6 & 59.2 \\

\bottomrule
\end{tabular}%
}
\caption{Ablation of the speech front-end and segment-level timestamps in \ourmodel{} across six audio-visual benchmarks.}
\label{tab:aba_av_benchmark_results}
\end{table}


\subsection{Ablation Study}
\label{sec:ablation}


In Table~\ref{tab:aba_av_benchmark_results}, we ablate the contributions of the frozen VLM, Whisper-based speech routing, and segment-level timestamps. We use Qwen2.5-VL-3B as the frozen backbone and evaluate across six audio-visual understanding benchmarks. The frozen VLM alone already provides strong video understanding, outperforming Qwen2.5-Omni on WorldSense and Video-Holmes. Adding Whisper supplies the missing spoken evidence while preserving this visual capability, improving performance across all benchmarks and surpassing Qwen2.5-Omni on five of six benchmarks. Removing segment-level timestamps reduces performance on AVUT-Gemini and Daily-Omni by \textbf{2.3} and \textbf{1.4} points, respectively, showing that timestamps provide targeted benefits for temporal audio-video alignment. Overall, the ablation isolates the role of each component: the frozen VLM contributes video reasoning, Whisper contributes speech understanding, and timestamps support temporal alignment.
\section{Conclusion}
\label{sec:conclusion}

Native omni training is costly and brittle, difficult to align reliably across noisy and temporally dense audio-visual signals, and may weaken capabilities already learned by the original VLM. So, \textit{Do we need to rebuild a native omni model every time a stronger VLM becomes available?} To investigate this question, we introduced \ourmodel{}, a training-free framework that routes confidence-filtered, timestamped speech transcripts through the language interface of a frozen VLM, without modifying its architecture or visual pathways. Across 56 benchmarks and 21 languages, our results address the three challenges. \textit{First}, \ourmodel{} removes backbone-specific audio training and re-alignment, enabling speech-centric omni capability to be transferred to stronger VLMs. \textit{Second}, language-level routing performs similarly to or better than native omni models when spoken content and temporal speech-video evidence are central, with strong results on audio-visual, audio-only, and multilingual speech tasks. \textit{Third}, freezing the backbone preserves stronger image/video understanding, visual grounding, coding, mathematical reasoning, medical knowledge, and other downstream capabilities than the corresponding native omni checkpoints in most comparisons. The remaining gap lies in non-speech acoustic understanding, including music, environmental sounds, vocal tone, and emotion, where native acoustic training remains important. These findings establish a practical direction for future omni models: combine external ASR-based speech routing with careful acoustic training while preserving the mature capabilities of the underlying VLM.

\section*{Acknowledgments}
This work was supported in part by the Google Gemini Academic Program through Google Cloud Credits.

{
    \small
    \bibliographystyle{ieeenat_fullname}
    \bibliography{main}
}

\clearpage
\appendix


\begin{table*}[t]
\centering
\small
\begin{tabular}{p{0.22\linewidth} p{0.70\linewidth}}
\toprule
\textbf{Capability axis} & \textbf{Benchmarks} \\
\midrule
AV omni understanding &
UnoBench and UnoBench-MC~\citep{chen2025unobench};
WorldSense~\citep{hong2025worldsense};
AV-Odyssey~\citep{gong2024avodyssey};
Video-Holmes~\citep{cheng2025videoHolmes};
FutureOmni~\citep{chen2026futureomni};
AVMeme-Full and AVMeme-Main~\citep{jiang2026avmeme};
AVUT-Human and AVUT-Gemini~\citep{yang2025avut};
Daily-Omni~\citep{zhou2025dailyomni};
AVHBench~\citep{kim2024avhbench} . \\
\midrule
Image/video understanding &
ChartQA~\citep{masry2022chartqa};
DocVQA~\citep{mathew2021docvqa};
InfoVQA~\citep{mathew2022infographicvqa};
OCRBench~\citep{liu2024ocrbench};
TextVQA~\citep{singh2019textvqa};
MMBench~\citep{liu2024mmbench};
MMStar~\citep{chen2024mmstar};
MME~\citep{fu2026mme};
LongVideoBench~\citep{wu2024longvideobench};
LVBench~\citep{wang2025lvbench};
MotionBench~\citep{hong2025motionbench};
VideoMME and VideoMME-Short~\citep{fu2024videomme}. 
EgoSchema~\citep{mangalam2023egoschema};
MVBench~\citep{li2024mvbench};
\\
\midrule
Audio-only understanding &
Audio Trivia \citep{joshi2017triviaqa};
Audio Web Questions \citep{berant2013semantic};
CoVoST2~\citep{wang2020covost};
FLEURS~\citep{conneau2022fleurs};
LibriSpeech~\citep{panayotov2015librispeech};
LiveSports3K~\citep{chen2025livecc};
MELD~\citep{poria2019meld};
MMAR-Bench~\citep{ma2025mmar};
Voice-CMMLU~\citep{li2024cmmlu};
VoiceBench~\citep{chen2026voicebench}. \\

\midrule
Multilingual speech understanding &
CoVoST2~\citep{wang2020covost} evaluated across 21 languages: Arabic, Catalan, Welsh, German, Spanish, Estonian, Persian, French, Indonesian, Italian, Japanese, Latvian, Mongolian, Dutch, Portuguese, Russian, Slovenian, Swedish, Tamil, Turkish, and Chinese. \\
\midrule
Coding and mathematical reasoning &
MBPP \cite{austin2021mbpp};
MBPP Sanitized \cite{austin2021mbpp};
HumanEval \cite{chen2021openai_humaneval};
GSM8K \cite{cobbe2021gsm8k};
MATH500 \cite{lightman2024let};
MathVerse~\citep{zhang2024mathverse};
MathVista~\citep{lu2024mathvista};
VideoMath~\citep{rasheed2025videomathqa}.\\
\midrule
Medical question answering &
MMMU-Med-test \cite{yue2024mmmu};
MMMU-Med-val \cite{yue2024mmmu};
MedFrameQA \cite{yu2025medframeqa};
MedMCQA \cite{pal2022medmcqa};
MedQA-USMLE \cite{jin2020medqa};
MedXQA-MM \cite{zuo2025medxpertqa};
MedXQA-Text \cite{zuo2025medxpertqa};
Medbullets-op4 \cite{chen2024medbulltes};
Medbullets-op5 \cite{chen2024medbulltes};
OmniMedVQA \cite{hu2024omnimedvqa};
PATH-VQA \cite{he2020pathvqa};
PMC-VQA \cite{zhang2023pmc};
PubMedQA \cite{jin2019pubmedqa};
SGPQA \cite{pteam2025supergpqascalingllmevaluation}. \\

\midrule
Grounding &
PixMo \cite{deitke2025molmo};
RefCoCo \cite{kazemzadeh2014refcoco};
PointArena \cite{cheng2025pointarena}. \\

\bottomrule
\end{tabular}
\caption{
Benchmark suite used in the core evaluation. Counting by the identity of the underlying benchmark dataset, the suite contains 56 distinct datasets: 10 AV omni, 14 image/video, 9 audio-only, 7 coding/math, 12 medical QA, and 4 grounding datasets. Subsets, data splits, and alternative evaluation variants of the same underlying dataset are counted once, and the 21 CoVoST2 language splits are not counted separately. LiveSports3K is evaluated only in the supplementary audio-only ablation and is excluded from this core total.
}
\label{tab:benchmark_suite_appendix}
\end{table*}

\begin{figure*}[t]
\centering
\begin{tcolorbox}[
    width=0.96\textwidth,
    colback=white,
    colframe=black!55,
    colbacktitle=blue!7,
    coltitle=black,
    title=\textbf{Generalized \ourmodel{} Prompt Template},
    fonttitle=\small\bfseries,
    boxrule=0.5pt,
    arc=2mm,
    outer arc=2mm,
    left=7pt,
    right=7pt,
    top=5pt,
    bottom=5pt,
    toptitle=4pt,
    bottomtitle=4pt
]
\small
\textbf{System:}\\
\texttt{<dataset-specific system instruction>}

\vspace{4pt}
\textbf{Audio file transcript:}\\
\texttt{Audio Language: <language> (<confidence>\%)}\\
\texttt{Whisper Transcript:}\\
\texttt{<separate-audio transcript>}

\vspace{4pt}
\textbf{Video audio transcript:}\\
\texttt{Audio Language: <language> (<confidence>\%)}\\
\texttt{Whisper Transcript:}\\
\texttt{<embedded-video transcript>}

\vspace{4pt}
\textbf{Visual input:}\\
\texttt{<image(s) or sampled video frames>}

\vspace{4pt}
\textbf{User:}\\
\texttt{<question, answer choices, and output instruction>}

\vspace{5pt}
\textit{When only one audio source is available, it is included under a
single \texttt{Audio Transcript:} field.}
\end{tcolorbox}
\caption{Generalized prompt structure used by the VLM+Whisper implementations. Separate and embedded video audio are independently labelled when both are available. With a single source, one \texttt{Audio Transcript:} block is used. Only available and confidence-filtered transcripts are included.}
\label{fig:prompt_template}
\end{figure*}

\section{Benchmark Suite}
\label{supp:app_benchmark_suite}

Table~\ref{tab:benchmark_suite_appendix} summarizes the core evaluation suite by capability axis. We compute the total according to the identity of the underlying benchmark dataset, rather than counting every subset, split, or evaluation variant separately. Accordingly, UnoBench and UnoBench-MC, AVMeme-Full and AVMeme-Main, VideoMME and VideoMME-Short, MBPP and MBPP Sanitized, MMMU-Med-test and MMMU-Med-val, and Medbullets-op4 and Medbullets-op5 are each counted once. CoVoST2 is also counted once, although it is used in both the audio-only and multilingual evaluations, and its 21 language splits are not counted as separate datasets. In contrast, AVUT-Human and AVUT-Gemini are counted separately because they contain different samples and use different annotation pipelines; MedXQA-MM and MedXQA-Text are counted separately because they evaluate different input modalities; and PixMo-Count and PixMo-Point are counted separately because they are distinct datasets. Under this protocol, the core evaluation contains 56 distinct benchmark datasets: 10 AV omni, 14 image/video, 9 audio-only, 7 coding/math, 12 medical QA, and 4 grounding datasets. LiveSports3K is used only in the supplementary audio-only ablation in Table~\ref{tab:aba_audio_benchmark_results} and is therefore excluded from the 56-dataset core total; including this ablation-only dataset, the paper reports results on 57 distinct datasets overall.

\section{Evaluation Protocol}
\label{supp:app_appendix_evaluation}

We use task-specific evaluation protocols according to the required output format. For multiple-choice tasks, performance is measured using exact-match accuracy between the predicted option and the ground-truth answer. For long-form generation tasks, we use GPT-5.6 (\texttt{GPT-5.6 Sol}) as an LLM-based evaluator. The evaluator receives both the reference response and the model prediction and determines whether the prediction is correct. For translation tasks, the evaluation considers lexical accuracy, preservation of meaning, and overall translation quality. For open-ended question answering, a prediction is considered correct when it is semantically consistent with the reference answer, even when the wording differs.

Each model is evaluated once on each benchmark. Model inference is performed using greedy decoding with the temperature set to zero. This deterministic decoding configuration ensures that repeated inference with the same model and input produces consistent outputs, subject to the determinism of the underlying software and hardware implementation.

All model inference experiments are conducted using AMD Instinct MI210 GPUs with 64\,GB of memory per GPU. To standardize the evaluation process, we developed \textbf{\textit{OmniEvalKit}}, a unified evaluation toolkit for multimodal and omni-modal models. OmniEvalKit provides a consistent interface for dataset loading, prompt construction, model inference, output parsing, and metric computation across multiple task formats. The complete toolkit will be released as open-source to support reproducibility and future evaluation.

\section{Additional Implementation Details}
\label{supp:app_implementation_details}

\subsection{Prompt Construction.}
\label{supp:app_promt_construction}
Following Eq.~\ref{eq:tfo_inference}, \ourmodel{} converts the confidence-filtered transcript $T_{\tau}(A)$ into a language-side audio context $C_A$. For speech-content reasoning, $C_A$ contains the retained spoken segments. For temporal audio-video reasoning, their timestamps are also preserved, providing anchors for relating spoken content to visual events. The final prompt is constructed as
\begin{equation}
    P =
    [\text{system}]
    \oplus
    [V]
    \oplus
    [C_A]
    \oplus
    [Q],
    \label{eq:tfo_prompt}
\end{equation}
where $[\text{system}]$ denotes the dataset-specific instruction, $[V]$ the visual input processed through the backbone's original visual pathway, $[C_A]$ the routed speech context, and $[Q]$ the user query, including answer choices and output constraints when applicable. Thus, speech conditions the VLM only through its language interface, while its visual pathway remains unchanged. If $T_{\tau}(A)=\emptyset$, the audio context is omitted and the model receives its original image/video-text prompt.

As illustrated in Figure~\ref{fig:prompt_template}, separate audio files and embedded video audio are represented by explicitly labeled blocks, such as \texttt{Audio file transcript:} and \texttt{Video audio transcript:}. When both sources are available, they are transcribed independently and both blocks are included. When only one source is available, it is inserted under a single \texttt{Audio Transcript:} field. These labels preserve the origin of each transcript within the language context.

\subsection{Audio Preprocessing.}
\label{supp:app_audio_preprocessing}
For \ourmodel{}, audio is converted to mono 16-kHz waveforms and transcribed deterministically using Whisper-large-v3-turbo in transcription mode with temperature zero. Language identification is performed before transcription, and segments below the confidence threshold of $0.65$ are discarded. Qwen-based implementations process long audio in 30-second chunks with a 3-second overlap. If no reliable speech is detected, the VLM receives only the original visual input and task prompt. Unless otherwise specified in the ablation studies, auxiliary audio models such as SenseVoice and MELLOW are disabled.


\begin{table*}[!t]
\centering
\small
\setlength{\tabcolsep}{3.5pt}
\renewcommand{\arraystretch}{0.95}
\resizebox{\textwidth}{!}{%
\begin{tabular}{@{}lccccccccccc@{}}
\toprule
\textbf{Method}
& \textbf{UnoBench}
& \textbf{Uno-MC}
& \textbf{WorldSense}
& \textbf{AV-Odyssey}
& \textbf{V-Holmes}
& \textbf{FutureOmni}
& \textbf{AVMeme-F}
& \textbf{AVMeme-M}
& \textbf{AVUT-H}
& \textbf{AVUT-G}
& \textbf{Daily-Omni} \\
\midrule

Qwen2.5-Omni
& \textbf{38.4} & 35.0 & 34.2 & 26.3 & 34.6
& 42.9 & \textbf{53.0} & \textbf{47.0}
& 55.0 & 55.5 & 53.7 \\

\midrule

Qwen2.5-VL
& 24.4 & 33.5 & 39.7 & 25.3 & 39.9
& 45.6 & 46.5 & 40.0 & 52.7 & 42.9 & 47.2 \\

\rowcolor{oursblue}
\hspace{0.4em}+ Whisper
& 38.0 & 35.1 & 40.3 & 26.8 & \textbf{42.7}
& 45.9 & 51.3 & 45.0 & 63.6 & \textbf{58.9} & \textbf{60.6} \\

\hspace{0.4em}+ Whisper (W/o TS)
& 38.0 & 35.1 & 40.1 & 26.2 & 42.5
& 45.9 & 49.6 & 44.1 & 62.9 & 56.6 & 59.2 \\

\hspace{0.4em}+ Whisper ($\tau=0.6$)
& 38.0 & 35.1 & -- & 26.3 & 42.7
& -- & 51.0 & 44.6 & 63.5 & 58.9 & 60.6 \\

\hspace{0.4em}+ Whisper ($\tau=0.75$)
& 38.0 & 34.3 & -- & 26.2 & 42.7
& -- & 51.0 & 44.6 & 63.6 & 58.9 & 60.6 \\

\hspace{0.4em}+ Whisper + SV
& 38.1 & \textbf{35.3} & \textbf{45.6} & 26.6 & 41.4
& 46.4 & 51.0 & 45.1 & \textbf{63.7} & 58.8 & 60.5 \\

\hspace{0.4em}+ Whisper + AF2
& 37.0 & 32.0 & 39.6 & 26.0 & 34.5
& 41.5 & 38.2 & 32.6 & 57.7 & 55.8 & 39.5 \\

\hspace{0.4em}+ Whisper + MEL
& 37.8 & 35.0 & 44.6 & 26.7 & 40.9
& 46.5 & 51.4 & 45.1 & 62.9 & 58.2 & 59.6 \\

\hspace{0.4em}+ Whisper + BEATs
& 33.2 & 32.4 & 37.8 & 20.3 & 42.6
& 44.2 & 49.5 & 43.4 & 60.7 & 56.6 & 56.8 \\

\hspace{0.4em}+ Whisper + CLAP
& 32.1 & 31.3 & 37.5 & 19.4 & 41.8
& 44.5 & 49.1 & 43.1 & 60.5 & 56.8 & 55.5 \\

\hspace{0.4em}+ Whisper + MEL + SV
& 37.9 & 35.1 & 45.4 & \textbf{26.9} & 40.9
& \textbf{47.0} & 51.0 & 45.3 & 62.7 & 58.7 & 59.6 \\

\bottomrule
\end{tabular}%
}
\caption{
\textbf{Audio-visual ablation on Qwen2.5-VL-3B.}
Whisper provides the main gains, timestamps benefit temporal reasoning, and auxiliary audio models offer no consistent improvement. Uno-MC denotes UnoBench-MC; AVMeme-F/M denote AVMeme-Full/Main; SV denotes SenseVoice; AF2 denotes AudioFlamingo2; MEL denotes MELLOW; and W/o TS denotes without timestamps.
}
\label{tab:aba_av_benchmark_results}
\end{table*}



\begin{table*}[!t]
\centering
\small
\setlength{\tabcolsep}{4.2pt}
\renewcommand{\arraystretch}{0.95}
\resizebox{\textwidth}{!}{%
\begin{tabular}{@{}lccccccccc@{}}
\toprule
\textbf{Method}
& \textbf{Audio Trivia}
& \textbf{Audio Web}
& \textbf{FLEURS}
& \textbf{LibriSpeech}
& \textbf{LiveSports3K}
& \textbf{MELD}
& \textbf{MMAR-Bench}
& \textbf{Voice-CMMLU}
& \textbf{VoiceBench} \\
\midrule

Qwen2.5-Omni
& 46.3 & 40.4 & 70.9 & \textbf{95.0}
& 15.1 & 53.4 & \textbf{55.4} & 55.6 & 66.0 \\

\midrule

\rowcolor{oursblue}
Qwen2.5-VL + Whisper
& 46.4 & \textbf{45.7} & \textbf{72.2}
& 94.7 & \textbf{20.9} & \textbf{54.4} & 49.7
& \textbf{58.9} & 66.3 \\

\hspace{0.4em}+ MELLOW
& 45.5 & 39.6 & 70.9 & 92.3
& 17.9 & 49.7 & 48.3 & 58.6 & 65.5 \\

\hspace{0.4em}+ SenseVoice
& \textbf{47.1} & 44.3 & 71.1 & 92.5
& 17.6 & 53.7 & 48.7 & 58.6 & \textbf{66.5} \\

\hspace{0.4em}+ MELLOW + SenseVoice
& 45.6 & 40.3 & 70.1 & 92.1
& 11.4 & 50.4 & 49.0 & 58.5 & 65.5 \\

\bottomrule
\end{tabular}%
}
\caption{
\textbf{Audio-only ablation on Qwen2.5-VL-3B.}
Whisper provides the strongest overall configuration, while auxiliary audio outputs offer no consistent benefit.
}
\label{tab:aba_audio_benchmark_results}
\end{table*}


\section{Additional Ablations}
\label{supp:app_additional_ablations}

\paragraph{Motivation and auxiliary audio models.}
The original Qwen2.5-VL already performs better than Qwen2.5-Omni on visually informative benchmarks such as WorldSense and Video-Holmes, suggesting that its visual reasoning should be preserved. To provide access to the spoken information in audio and video, we use Whisper~\cite{whisper}. This improves all six benchmarks over the original VLM and surpasses Qwen2.5-Omni on five, showing that speech transcription complements the preserved visual backbone effectively. But, from the Tables \ref{tab:aba_av_benchmark_results} and \ref{tab:aba_audio_benchmark_results}, we observed that AVMeme and MMAR-Bench still not overcome the omni version of the model, revealing that transcription still misses emotion, vocal attributes, music, environmental sounds, and other non-speech acoustic events. We therefore ask: \textbf{\textit{can these missing cues be recovered using external audio models and injected through the same language interface?}} To examine this, we append textual outputs from complementary models to the Whisper transcript. SenseVoice~\cite{an2024funaudiollm} is used to capture speech emotion and audio events; MELLOW~\cite{deshmukh2026mellow} to reason over acoustic scenes and events; AudioFlamingo2 (AF2)~\cite{ghosh2025audioflamingo2} to describe general and long-form audio; BEATs~\cite{chen2022beats} to identify semantic sound events; and CLAP~\cite{elizalde2023clap} to provide open-vocabulary audio concepts through audio-text alignment. The VLM remains unchanged in all configurations.

\paragraph{Audio-visual ablation.}
Table~\ref{tab:aba_av_benchmark_results} evaluates how speech transcripts, timestamps, and auxiliary acoustic descriptions affect audio-visual understanding. To align spoken content with visual events, timestamps are retained in the Whisper transcript. Removing them causes the largest drops on AVUT-Gemini (\textbf{-2.3}), AVMeme-Full (\textbf{-1.7}), and Daily-Omni (\textbf{-1.4}), confirming their importance for temporally grounded reasoning. To capture the non-speech information missing from Whisper, several auxiliary audio models are then added. SenseVoice improves WorldSense by \textbf{+5.3} points, but its largest regressions occur on Video-Holmes (\textbf{-1.3}), AVMeme-Full (\textbf{-0.3}), and AV-Odyssey (\textbf{-0.2}). MELLOW similarly improves WorldSense by \textbf{+4.3}, but reduces Video-Holmes by \textbf{1.8}, Daily-Omni by \textbf{1.0}, and both AVUT benchmarks by \textbf{0.7}. AF2 produces substantially larger drops, including \textbf{21.1} on Daily-Omni, \textbf{13.1} on AVMeme-Full, and \textbf{12.4} on AVMeme-Main. BEATs and CLAP also degrade several tasks, particularly AV-Odyssey, UnoBench, and Daily-Omni. Combining MELLOW and SenseVoice retains a strong WorldSense gain (\textbf{+5.1}) but again fails to generalize. Overall, auxiliary descriptions occasionally help individual benchmarks, but none consistently outperforms timestamped Whisper.

\paragraph{Audio-only ablation.}
Table~\ref{tab:aba_audio_benchmark_results} examines whether auxiliary audio models can address the broader acoustic information missing from Whisper when no visual evidence is available. Whisper alone provides the strongest overall configuration, outperforming Qwen2.5-Omni on eight of ten benchmarks, including Audio Web (\textbf{+5.3}), CoVoST2 (\textbf{+5.4}), LiveSports3K (\textbf{+5.8}), and Voice-CMMLU (\textbf{+3.3}). However, its \textbf{-5.7}-point gap on MMAR-Bench confirms that transcription remains insufficient for mixed and non-speech acoustic reasoning. To address this limitation, SenseVoice and MELLOW are added individually and jointly. SenseVoice slightly improves Audio Trivia (\textbf{+0.7}) and VoiceBench (\textbf{+0.2}) over Whisper, but reduces LiveSports3K by \textbf{3.3}, LibriSpeech by \textbf{2.2}, and Audio Web by \textbf{1.4} points. MELLOW produces larger declines on CoVoST2 (\textbf{-9.2}), Audio Web (\textbf{-6.1}), and MELD (\textbf{-4.7}), while their combination further lowers CoVoST2 by \textbf{9.7}, LiveSports3K by \textbf{9.5}, and Audio Web by \textbf{5.4} points. These results suggest that auxiliary acoustic outputs may benefit isolated tasks, but simple textual concatenation does not consistently recover the missing non-speech information.

\begin{table*}[!t]
\centering
\setlength{\tabcolsep}{2.2pt}
\renewcommand{\arraystretch}{1.12}
\resizebox{1.0\textwidth}{!}{%
\begin{tabular}{@{}lcccccccccccccccccccccc@{}}
\toprule
\textbf{Model}
& \textbf{ar}
& \textbf{ca}
& \textbf{cy}
& \textbf{de}
& \textbf{es}
& \textbf{et}
& \textbf{fa}
& \textbf{fr}
& \textbf{id}
& \textbf{it}
& \textbf{ja}
& \textbf{lv}
& \textbf{mn}
& \textbf{nl}
& \textbf{pt}
& \textbf{ru}
& \textbf{sl}
& \textbf{sv}
& \textbf{ta}
& \textbf{tr}
& \textbf{zh}
& \textbf{Avg.} \\
\midrule

Qwen2.5-Omni-3B
& 55.6 & 41.3 & 7.6 & 61.1 & 64.5 & 12.3 & 9.0 & 62.9 & 61.3 & 62.1 & 47.4 & 6.9 & 3.1 & 63.3 & 73.5 & 69.3 & 14.9 & 26.2 & 6.6 & 34.2 & 51.9 & 39.8 \\

\rowcolor{oursblue}
Qwen2.5-VL-\ourmodel-3B
& 48.0 & 47.2 & 13.5 & 59.7 & 62.5 & 28.6 & 25.1 & 59.5 & 63.6 & 58.4 & 46.2 & 21.6 & 7.2 & 59.2 & 69.1 & 63.8 & 29.7 & 46.1 & 9.3 & 45.6 & 42.8 & \textbf{43.2} \\

Qwen2.5-Omni-7B
& 59.2 & 40.8 & 7.4 & 62.1 & 65.2 & 13.6 & 7.6 & 64.0 & 64.1 & 62.3 & 28.6 & 7.0 & 4.1 & 62.6 & 70.9 & 68.9 & 15.4 & 12.8 & 4.8 & 18.4 & 49.3 & 37.6 \\

\rowcolor{oursblue}
Qwen2.5-VL-\ourmodel-7B
& 45.2 & 49.0 & 18.4 & 60.2 & 61.2 & 35.4 & 28.3 & 59.6 & 64.9 & 58.3 & 47.5 & 26.6 & 5.2 & 61.5 & 71.2 & 65.8 & 32.3 & 49.7 & 12.0 & 49.1 & 43.6 & \textbf{45.0} \\

\midrule
MiniCPM4.5-O-9B
& 25.6 & 34.8 & 8.1 & 50.4 & 58.1 & 18.9 & 16.1 & 52.6 & 47.6 & 51.0 & 29.4 & 7.0 & 7.4 & 46.1 & 56.5 & 57.9 & 17.3 & 26.5 & 7.9 & 38.4 & 48.2 & 33.6 \\

\rowcolor{oursblue}
MiniCPM4.5-V-\ourmodel-9B
& 45.9 & 49.3 & 20.9 & 61.0 & 63.0 & 36.9 & 28.4 & 59.7 & 66.6 & 60.0 & 49.9 & 28.4 & 7.8 & 62.2 & 71.3 & 66.6 & 34.7 & 51.3 & 15.2 & 49.6 & 44.1 & \textbf{46.3} \\

\midrule
OmniVinci-9B
& 28.7 & 42.0 & 8.1 & 58.4 & 63.1 & 14.4 & 8.8 & 60.8 & 32.9 & 59.2 & 42.5 & 8.0 & 8.1 & 50.8 & 67.6 & 60.6 & 13.9 & 13.5 & 7.5 & 17.6 & 41.8 & 33.7 \\

\rowcolor{oursblue}
NVILA-8B-VL-\ourmodel
& 32.5 & 42.5 & 13.2 & 58.7 & 60.8 & 26.2 & 18.0 & 58.1 & 60.5 & 57.3 & 46.1 & 18.7 & 8.1 & 57.4 & 68.7 & 62.9 & 25.4 & 41.4 & 7.7 & 38.1 & 40.9 & \textbf{40.2} \\

\midrule
Qwen3-Omni-30B-A3B
& 68.8 & 42.4 & 8.4 & 65.6 & 66.8 & 22.6 & 27.2 & 65.4 & 72.9 & 65.6 & 54.7 & 9.8 & 7.4 & 67.4 & 74.8 & 71.2 & 28.2 & 50.1 & 10.7 & 58.1 & 51.8 & 47.1 \\

\rowcolor{oursblue}
Qwen3-VL-30B-A3B-\ourmodel
& 49.3 & 52.7 & 24.9 & 63.7 & 64.9 & 43.1 & 32.1 & 61.8 & 70.1 & 62.5 & 52.8 & 35.0 & 6.5 & 65.2 & 73.8 & 70.8 & 39.4 & 54.1 & 17.5 & 53.8 & 48.2 & \textbf{49.6} \\

\bottomrule
\end{tabular}%
}

\caption{\textbf{Multilingual speech translation on CoVoST2.}
We report BLEU-1 scores across 21 languages, where higher values indicate better translation quality. The final average is the unweighted macro-average of the 21 language-level BLEU-1 scores.}
\label{supp:app_multilingual}
\end{table*}

\paragraph{Language-confidence threshold ablation.}
We study the minimum speech detection confidence threshold $\tau$ applied before transcription. When the detected confidence is below $\tau$, Whisper transcription is skipped. As shown in Table~\ref{tab:aba_av_benchmark_results}, $\tau=0.60$ and $\tau=0.65$ perform similarly across the reported benchmarks. We use $\tau=0.65$ by default because, in our observations, it more reliably suppresses spurious outputs from music or non-speech regions, such as \eighthnote, \twonotes, or \texttt{[Music]}. Increasing the threshold to $\tau=0.75$ applies stricter filtering and may reject uncertain but informative speech, reducing UnoBench-MC from \textbf{35.1} to \textbf{34.3} and AV-Odyssey from \textbf{26.8} to \textbf{26.2}. These results suggest that moderate confidence filtering reduces unreliable ASR outputs without sacrificing useful spoken evidence, whereas excessive filtering can remove informative speech.

Overall, these ablations show that the gains of \ourmodel{} primarily come from reliable speech transcription and temporal anchoring. Although auxiliary audio models expose additional acoustic information, converting their outputs into text often introduces noisy or irrelevant context and does not reliably address the non-speech limitation. Richer acoustic representations may therefore require a more suitable integration mechanism than direct textual concatenation.

\section{Benchmark-Specific Evaluation Metrics and Aggregation}
\label{supp:benchmark_metrics_aggregation}

\paragraph{Evaluation of Table~\ref{tab:av_benchmark_results}.}
Table~\ref{tab:av_benchmark_results} contains 11 result rows corresponding to nine underlying audio-visual benchmarks. All closed-form questions are evaluated using exact option accuracy, where invalid, missing, or conflicting predictions are counted as incorrect. UnoBench contains both multiple-choice and open-ended questions: multiple-choice samples are evaluated using exact option accuracy, while open-ended samples are evaluated using binary semantic correctness with \texttt{GPT-5.6 Sol}, following Sec.~\ref{supp:app_appendix_evaluation}. The reported UnoBench score is the percentage of correct predictions across both question types, whereas UnoBench-MC reports the multiple-choice subset separately. Similarly, AVMeme-Main is a diagnostic subset of AVMeme-Full. Therefore, UnoBench-MC and AVMeme-Main are shown for detailed analysis but are excluded from the macro-average. AVUT-Human and AVUT-Gemini contain different samples and annotation settings and are therefore treated as two distinct benchmarks. The reported macro-average assigns one contribution to each of the nine underlying benchmarks, preventing datasets with additional subsets or evaluation variants from receiving extra weight. All scores are percentages, and higher values indicate better performance.

\paragraph{Evaluation of Table~\ref{tab:audio_benchmark_results}.}
All results in Table~\ref{tab:audio_benchmark_results} are reported using higher-is-better metrics. Audio Trivia and Audio Web Questions already provide released spoken-question audio. We use the released audio directly and do not synthesize evaluation inputs. We evaluate them as audio benchmarks by presenting only the spoken question to the model; the corresponding text question stored in the dataset JSON is not used as input. Predictions are evaluated using binary semantic correctness with \texttt{GPT-5.6 Sol}, following the LLM-based evaluation protocol described in Sec.~\ref{supp:app_appendix_evaluation}.
The evaluator receives the reference answer and the model prediction and determines whether the prediction is semantically consistent with the reference. For LibriSpeech and FLEURS, we report LLM-judged transcription accuracy rather than WER or CER. The evaluator is explicitly instructed to assign a correct label when the prediction preserves at least $80\%$ of the reference transcription's spoken content, based on semantic and lexical coverage. Differences in capitalization, punctuation, spacing, or minor wording are ignored when they do not alter the recognized content. The reported LibriSpeech and FLEURS scores correspond to the percentage of samples judged correct. MELD is evaluated using weighted F1 for emotion recognition, while MMAR-Bench, Voice-CMMLU, and VoiceBench are evaluated using exact-match accuracy. CoVoST2 is evaluated using the thresholded translation accuracy defined in the following paragraph.

\begin{figure*}[t]
    \centering
    \begin{minipage}{0.55\linewidth}
        \centering
        \includegraphics[width=\linewidth]{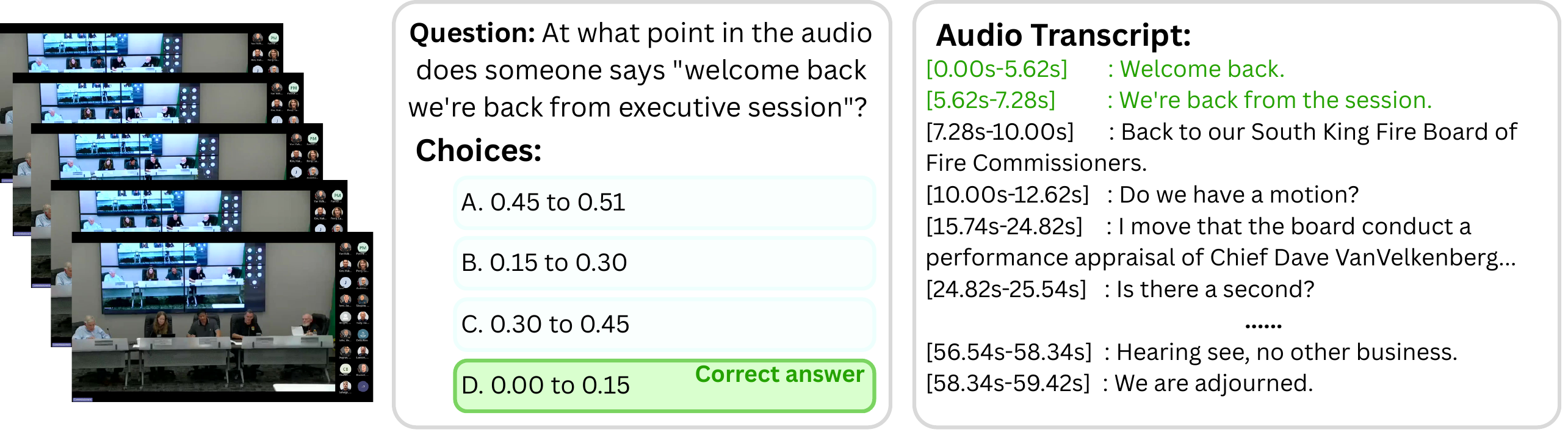}
    \end{minipage}
    \hfill
    \begin{minipage}{0.44\linewidth}
        \centering
        \includegraphics[width=\linewidth]{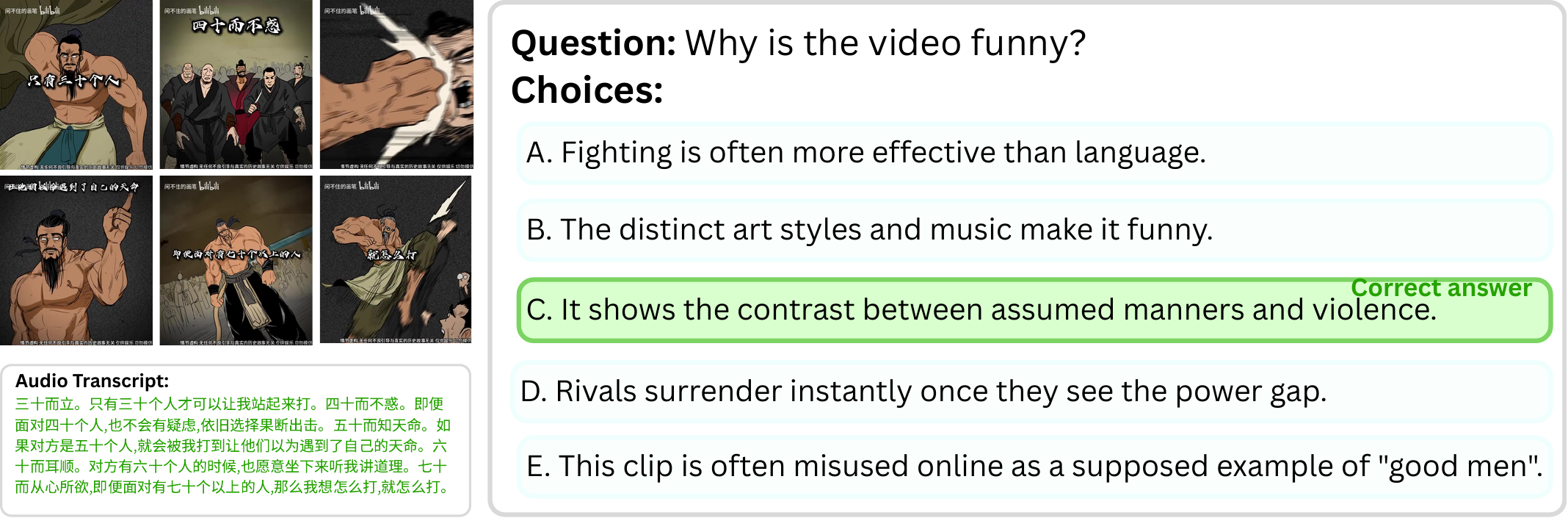}
    \end{minipage}
    \caption{
    \textbf{Qualitative examples of \ourmodel{}.}
    \textit{Left:} Timestamped transcripts provide explicit temporal anchors for locating a queried utterance and selecting the correct interval.
    \textit{Right:} Chinese speech is transcribed and combined with visual evidence and an English question, enabling multilingual audio-visual reasoning.
    }
    \label{fig:qualitative_results}
\end{figure*}

\paragraph{Evaluation of Tables~\ref{tab:multilingual} and~\ref{supp:app_multilingual}.}
Tables~\ref{tab:multilingual} and~\ref{supp:app_multilingual} evaluate multilingual speech translation on CoVoST2 across 21 languages using two complementary metrics. In Table~\ref{tab:multilingual}, we report thresholded translation accuracy. The prediction and reference are normalized using Unicode normalization, case folding, punctuation removal, and whitespace normalization. A prediction is counted as correct when the normalized prediction exactly matches the normalized reference or achieves a ROUGE-L F1 score of at least $0.40$. Each language score is the percentage of correctly translated samples, and the reported average is the unweighted macro-average across the 21 languages. Table~\ref{supp:app_multilingual} reports mean sentence-level BLEU-1, computed directly between each generated translation and its reference without applying the ROUGE-L correctness threshold. Importantly, BLEU-1 reproduces the same overall finding as thresholded accuracy: \ourmodel{} improves the average across all five model settings by \textbf{+3.4}, \textbf{+7.4}, \textbf{+12.7}, \textbf{+6.5}, and \textbf{+2.5} points, respectively, with the largest gain again obtained on MiniCPM4.5 (\textbf{33.6$\rightarrow$46.3}). Large improvements also remain concentrated in languages where the native Omni models are weak, including Latvian (\textbf{+21.4}) and Swedish (\textbf{+24.8}) on MiniCPM4.5, and Swedish (\textbf{+36.9}) and Turkish (\textbf{+30.7}) on Qwen2.5-7B. Thus, both the thresholded correctness metric and the lexical-overlap metric consistently support the conclusion that the ASR front-end transfers broad multilingual speech coverage to frozen VLMs.

\paragraph{Evaluation of Table~\ref{tab:iv_benchmark_results}.}
All scores in Table~\ref{tab:iv_benchmark_results} are reported as percentages, with higher values indicating better performance. ChartQA, DocVQA, InfoVQA, OCRBench, and TextVQA are evaluated using LLM-judged semantic correctness, following Sec.~\ref{supp:app_appendix_evaluation}. MMBench, MMStar, and MME are evaluated using exact programmatic accuracy. All video benchmarks, including LongVideoBench, LVBench, MotionBench, VideoMME, VideoMME-S, EgoSchema, and MVBench, are evaluated using exact multiple-choice accuracy.

\paragraph{Evaluation of Table~\ref{tab:coding_math}.}
For MBPP, MBPP-Sanitized, and HumanEval, we report
\emph{judge-based code-solution correctness} rather than the official
execution-based pass@1 metric. The generated responses may contain explanatory
text, intermediate reasoning, and code blocks that are not directly executable
without additional response parsing. We therefore use the fixed LLM-based
evaluation protocol described in
Sec.~\ref{supp:app_appendix_evaluation}. For each sample, the evaluator receives
the programming problem, reference solution, and generated response, and assigns
a binary correct or incorrect label based on whether the response specifies a
complete and logically correct solution to the requested task. These scores
measure semantic and algorithmic solution correctness and should not be
interpreted as evidence that the generated programs compile or pass the official
hidden unit tests.

GSM8K, MATH500, and MathVista are also evaluated using binary LLM-judge
correctness because model responses may include reasoning before the final
answer. MathVerse and VideoMath are evaluated using exact option-match accuracy.
All reported values are higher-is-better percentages. We report coding and
mathematical reasoning results separately and do not present the judge-based
coding scores as official execution-based pass@1 results.

\paragraph{Evaluation of Table~\ref{tab:all_medical_benchmark}.}
We evaluate all medical question-answering benchmarks in Table~\ref{tab:all_medical_benchmark} using the standardized evaluation pipeline provided by MedEvalKit\footnote{\url{https://github.com/alibaba-damo-academy/MedEvalKit}}. For multiple-choice benchmarks, the predicted answer option is extracted and compared exactly with the ground-truth option. For open-ended medical VQA benchmarks, predictions are evaluated using the benchmark-specific answer normalization and correctness rules implemented in MedEvalKit. All results are reported as accuracy percentages, with higher values indicating better performance.

CosyVoice3 is used only for optional spoken response generation and is not used to construct or modify any evaluation audio. Therefore, its speakers and accents do not affect the reported benchmark results. 
For reproducibility, the supplementary material includes the complete evaluation codebase, including dataset loaders, prompt templates, output parsers, metric implementations, LLM-judge prompts and configurations, and scripts for reproducing all reported benchmark scores.

\section{Qualitative Results}
\label{supp:app_qual_results}

Figure~\ref{fig:qualitative_results} presents two representative examples showing how the audio-to-language interface supports temporal grounding and multilingual audio-visual reasoning.

\textbf{Temporal grounding.}
In the left example, Whisper produces a segmented transcript with explicit timestamps. By matching the queried utterance, ``welcome back we're back from executive session,'' to the corresponding transcript interval, the VLM correctly selects the answer covering the beginning of the audio.

\textbf{Multilingual understanding.}
In the right example, the audio is in Chinese while the question is in English. Whisper preserves the Chinese speech as textual evidence, which the VLM combines with the visual context to correctly identify that the humor arises from the contrast between polite philosophical expressions and violent actions. This demonstrates cross-lingual audio-visual reasoning without native audio-language training.

\end{document}